\documentclass[twocolumn,aps,showpacs]{revtex4}
\usepackage{graphicx}
\usepackage{xcolor}
\usepackage{dcolumn}
\usepackage{bm}
\usepackage{epsf}
\usepackage{epsfig}
\usepackage{amsmath,amsfonts,amsthm, amssymb,latexsym}
\usepackage{enumerate}

\newcommand{\beq}{\begin{equation}}
\newcommand{\eeq}{\end{equation}}
\newcommand{\bcn}{\begin{center}}
\newcommand{\ecn}{\end{center}}

\newcommand{\lsim}{\lower0.5ex\hbox{$\; \buildrel < \over \sim \;$}}

\newcommand{\msun}{M_{\odot}}

\begin{document}

\title{Proton superconductivity and the masses of neutron stars}

\date{\today}

\author{Rodrigo Negreiros} \email{negreiros@if.uff.br}
\affiliation{Instituto de F{\'{i}}sica, Universidade Federal Fluminense, Av.\ Gal.\
Milton Tavares S/N, Niter{\'{o}}i, RJ, Brazil}

\author{Stefan Schramm} \email{schramm@th.physik.uni-frankfurt.de}
\affiliation{FIAS, Goethe University, Ruth Moufang Str.\ 1,
        60438 Frankfurt, Germany}

\author{Fridolin Weber} \email{fweber@mail.sdsu.edu}
\affiliation{{Department of Physics, San Diego State University, 5500
  Campanile Drive, San Diego, CA 92182}, \\
{Center for Astrophysics and Space Sciences, 
University of California, San Diego, La Jolla, CA 92093, USA
}}

\begin{abstract}
The unexpected temperature evolution of the neutron star in the
Cassiopeia A supernova remnant (Cas A, for short) has renewed
tremendous interest in the cooling mechanisms of neutron stars. In
particular, the formation of superconducting protons and superfluid
neutrons deep inside the  cores of neutron stars have become focal
points of the discussion. The purpose of this letter is to add a new
aspect to this discussion, which focuses on the connection between
proton superconductivity and the masses of neutron stars.
  
Assuming (as is currently the case) that the temperature evolution of
Cas A is largely controlled by superconducting protons, we study a
series of phenomenological proton-pairing models to determine how deep
into the stellar core superconducting protons actually penetrate.
This allows us to establish a heretofore unknown relationship between
the mass of the neutron star in Cas A and the penetration depth of the
superconducting proton phase. This relationship can be used to either
predict the depth of the superconducting proton phase, or, conversely,
determine the mass of Cas A from a reliable calculation of the size of
the proton superconducting phase in superdense neutron star matter.
We emphasize that the strategy outlined in this paper can be applied
to any other neutron star of similar age, whose temperature  might be
reliably monitored over a several years period. High-mass neutron
stars, such as the recently discovered neutron stars J1614-2230 ($1.97
\pm 0.04\, \msun$) and J0348+0432 ($2.01 \pm 0.04 \, \msun$), appear
particularly appealing as a significant fraction of the protons in
their cores may be superconducting.
\end{abstract}


\maketitle

{\it Introduction.}  One of the most prominent ways of probing the
poorly known core compositions and inner structures of neutron stars
consists in studying the thermal evolution of such objects
theoretically
\cite{Tsuruta1965,Maxwell1979,Schaab1996,Page2004,Page2006,Page2009,
  Blaschke2000,Grigorian2005,Blaschke2006,Page2011,Yakovlev2011,
  Voskresensky1997,Shternin2011,Shovkovy2002,Prakash1992,Lattimer1994,Gusakov2005,
  Horvath1991,Weber2007b,Negreiros2010,
  Niebergal,Alford2005a,Negreiros2012a} and then comparing the outcome
with the ever-increasing body of observed data on neutron star
temperatures
\cite{Page2004,Page2006,Page2011,Yakovlev2011,Heinke2010}. The quality
of observed neutron star temperatures has culminated in the recent
temperature data obtained for the compact object in the supernova
remnant Cassiopeia A over a 10-year period \cite{Heinke2010}. The
observed temperature evolution of Cas A exhibited a steeper
temperature drop than originally thought possible for a star of that
age ($\sim 300$ years).  Recent independent studies have shown,
however, that the observed behavior can be explained if the protons
form a $^1S_0$ superconductor with a pairing gap that is large and
non-vanishing throughout the entire stellar core, and the neutrons in
the core form a $^3P_2$ superfluid with a critical temperature  of
$5\times 10^9$~K \cite{Yakovlev2011,Page2011}.  The relatively sharp
decline of the temperature of Cas A is then caused by the emission of
neutrinos from neutron pair-breaking-formation processes
\cite{Yakovlev2011,Page2011}. It has also been shown that in-medium
effects play a very important role for the temperature evolution of
the neutron star in Cas A \cite{Blaschke2011}, and that the late onset
of the direct Urca process, driven by the changing core compression
during stellar spin-down, may provide an explanation of the observed
data as well \cite{Negreiros2013}.

Building on earlier work \cite{Yakovlev2011,Page2011}, here we too
assume that the neutrons in the core of the neutron star in Cas A have
formed a $^3P_2$ neutron superfluid. In contrast to the aforementioned
work, however, we treat proton pairing in neutron stars
\cite{Page2011b} on a more general footing by considering a variety of
different proton superconducting profiles and  critical temperatures.  As
shown
in this letter, this allows us to establish an important constraint on
the core density (core penetration depth) up to
which protons need to be superconducting so that agreement with the
temperature data of Cas A is achieved.  As in
\cite{Yakovlev2011,Page2011} we assume that the protons in the stellar
core are paired in the $^1S_0$ state since the first moments of the
star's evolution, which implies a critical temperature of $\sim
10^{9-10} K$ for proton superconductivity. The proton pairing is
necessary as it suppresses both the (density dependent) direct Urca
process among nucleons, $n \rightarrow p + e + \bar \nu_e$, as well as
the modified Urca process, $n + n \rightarrow n + p + e + \bar\nu_e$,
which would otherwise  spoil the agreement with the Cas A data.

{\it Proton Superconductivity Depth.}  We assume that the matter in
the core of a neutron star is well described by the
Akmall-Pandharipande-Ravenhall (APR) model for the nuclear equation of
state \cite{Akmal1998,Heiselberg2000}. We shall use a parametrized
version of this equation of state, as described in
\cite{Heiselberg2000}. The choice of parameters used in our study is
the same as in \cite{Negreiros2013}, such that the direct Urca process is
allowed for stars with masses greater than $1.4\, \msun$.
Following \cite{Yakovlev2011,Page2011} we allow the neutrons in the
core of the neutron star to form pairs in spin-triplet states
($^3P_2$).  As discussed in \cite{Yakovlev2011} neutrons in the crust
are likely to form spin-singlet ($^1S_0$) pairs, but these have
little effect on the thermal evolution. For the neutron triplet pairs
in the core we use a phenomenological model, as described in
\cite{Yakovlev2001a}.  The values of the superfluid gap, suppression
factors for neutrino emissivities, and the luminosity of the
pair-breaking-formation processes were also taken from
\cite{Yakovlev2001a}. 

To determine how deep superconducting protons actually penetrate into
the core of the neutron star in Cas A, we start our cooling
simulations for a superconducting proton phase that penetrates deeply
into the center of star.  i.e. in this case all the protons are paired and
have a very high critical temperature, meaning that the star begins its thermal
evolution with all protons paired up. As already known
from the work of \cite{Page2011,Yakovlev2011}, this assumption leads
to an agreement with the temperature data observed for Cas A. We then
repeat the cooling simulation, but for incrementally reduced depths of
the proton superconducting phase. This scheme is repeated until
agreement with the Cas A temperatures is no longer achieved.  The
cooling simulations are carried out for neutron stars with masses of
$1.4\,\msun$, $1.6\, \msun$, $1.7\, \msun$, and $1.8\,\msun$.  As in
\cite{Page2011}, all our cooling simulations are for an accreted
neutron star envelope. The value chosen for the accreted mass is
$\Delta M/M = 5.5\times 10^{-13}$.

\begin{figure}
 \centering
 \vspace{1.0cm}
 \includegraphics[width=8.cm]{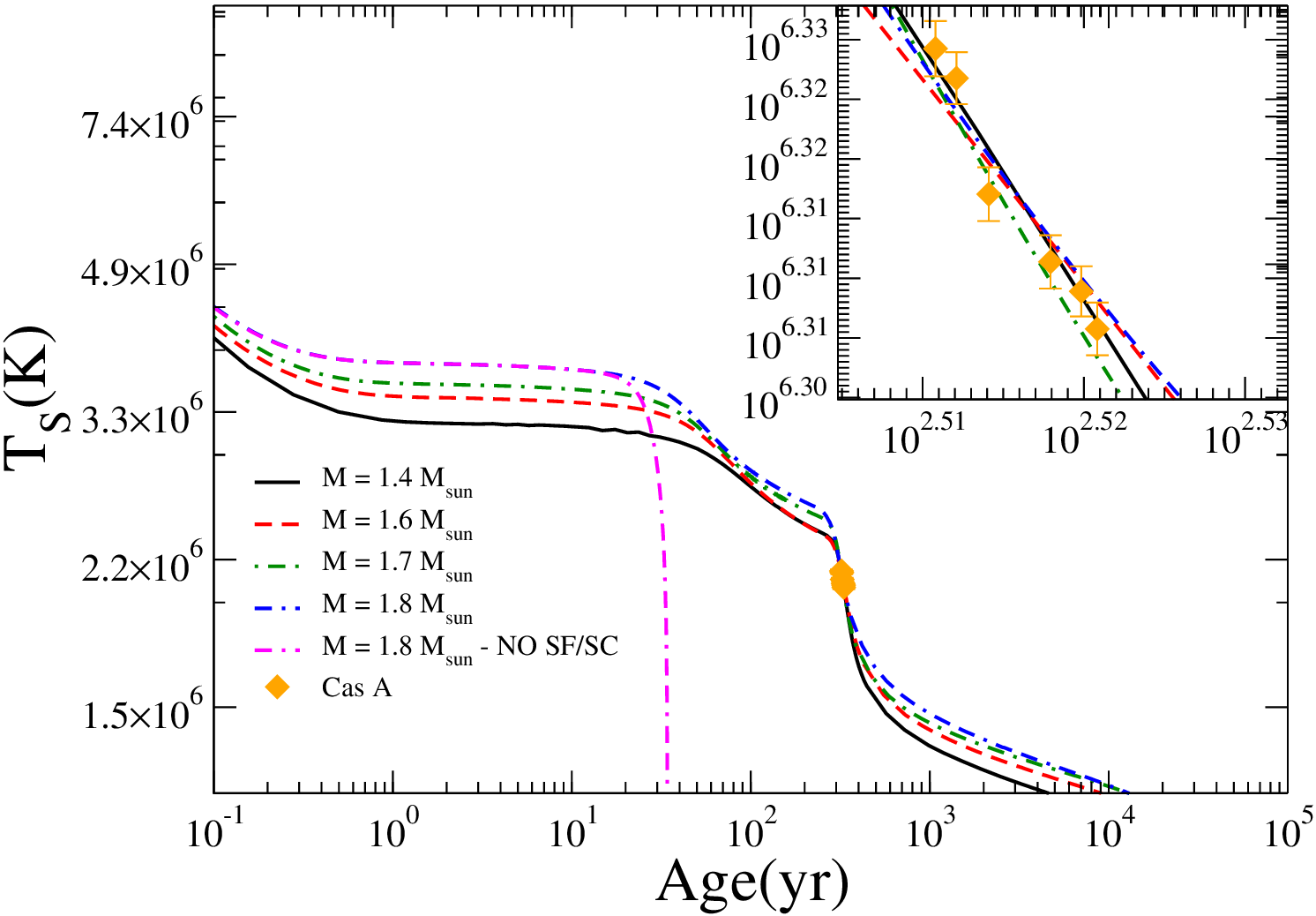}
 \caption{\label{cool_R}(Color online) Thermal evolution of neutron
   stars with different gravitational masses. The proton
   superconductivity depth inside of each star has been chosen such
   that agreement with the Cas A data is achieved. (See text for more
   details.)}
 \end{figure}

In Fig.\ \ref{cool_R}, we shown the temperature evolutions for our
sample neutron stars, each one possessing a superconducting proton 
core of proper size so that agreement with the Cas A data is achieved.
The insert in the upper right corner is an enlargements of the
temperature data of Cas A observe over a 10-year time period. Also
show in Fig.\ \ref{cool_R} is the temperature evolution of a $1.8\,
\msun$ neutron star when all neutrons and protons in the core are
unpaired. The latter badly fails to describe the Cas A data.

\begin{figure}
 \centering
 \vspace{1.0cm}
 \includegraphics[width=8.cm]{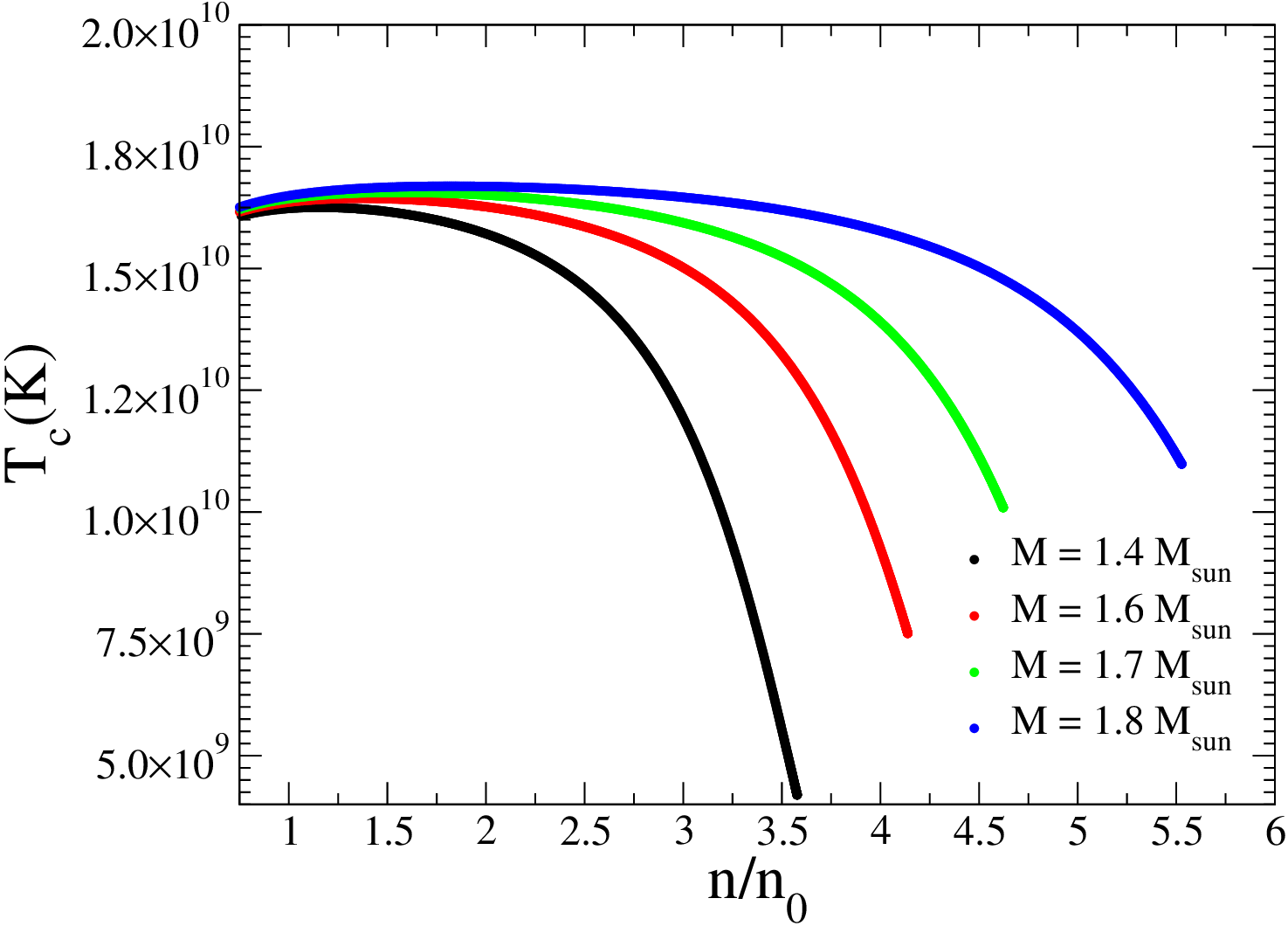}
 \caption{\label{pSC_prof}(Color online) Critical temperatures, $T_c$,
   of superconducting $^1S_0$ protons as a function of baryon number
   density, $n$, for the stellar models of Fig.\ \ref{cool_R}.
   ($n_0=0.16~{\rm fm}^{-3}$ denotes the baryon number density of
   ordinary nuclear matter.)}
     \end{figure}

\begin{figure}
 \centering
 \vspace{1.0cm}
 \includegraphics[width=8.cm]{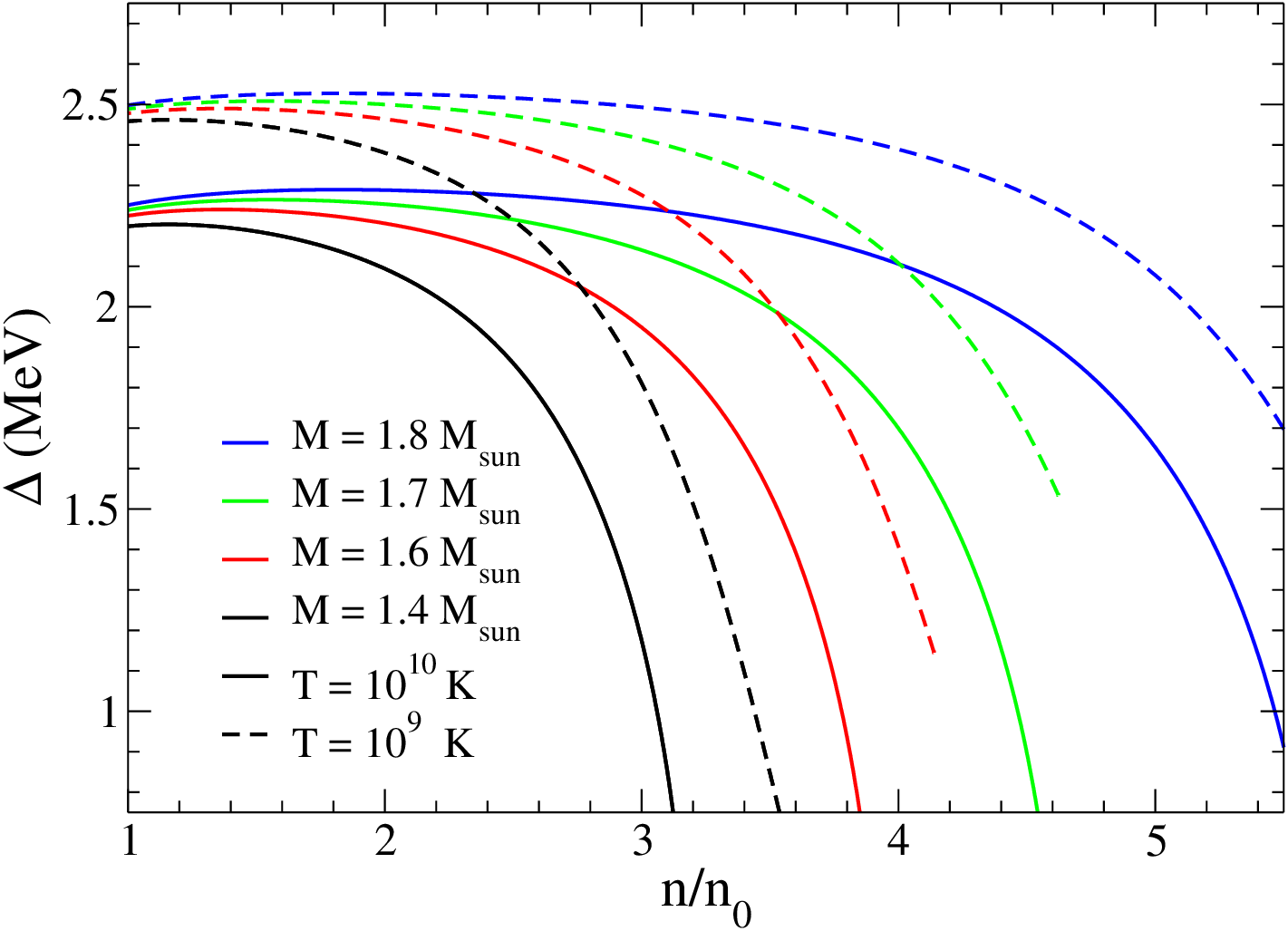}
 \caption{\label{GAP}(Color online) Gaps of superconducting $^1S_0$
   protons for the neutron star models shown in
   Fig.\ \ref{pSC_prof}. Solid lines denote a temperature of $10^{10}$
   K, dashed lines correspond to a temperature of $10^9$ K.}
     \end{figure}

\begin{figure}
 \centering
 \vspace{1.0cm}
 \includegraphics[width=8.cm]{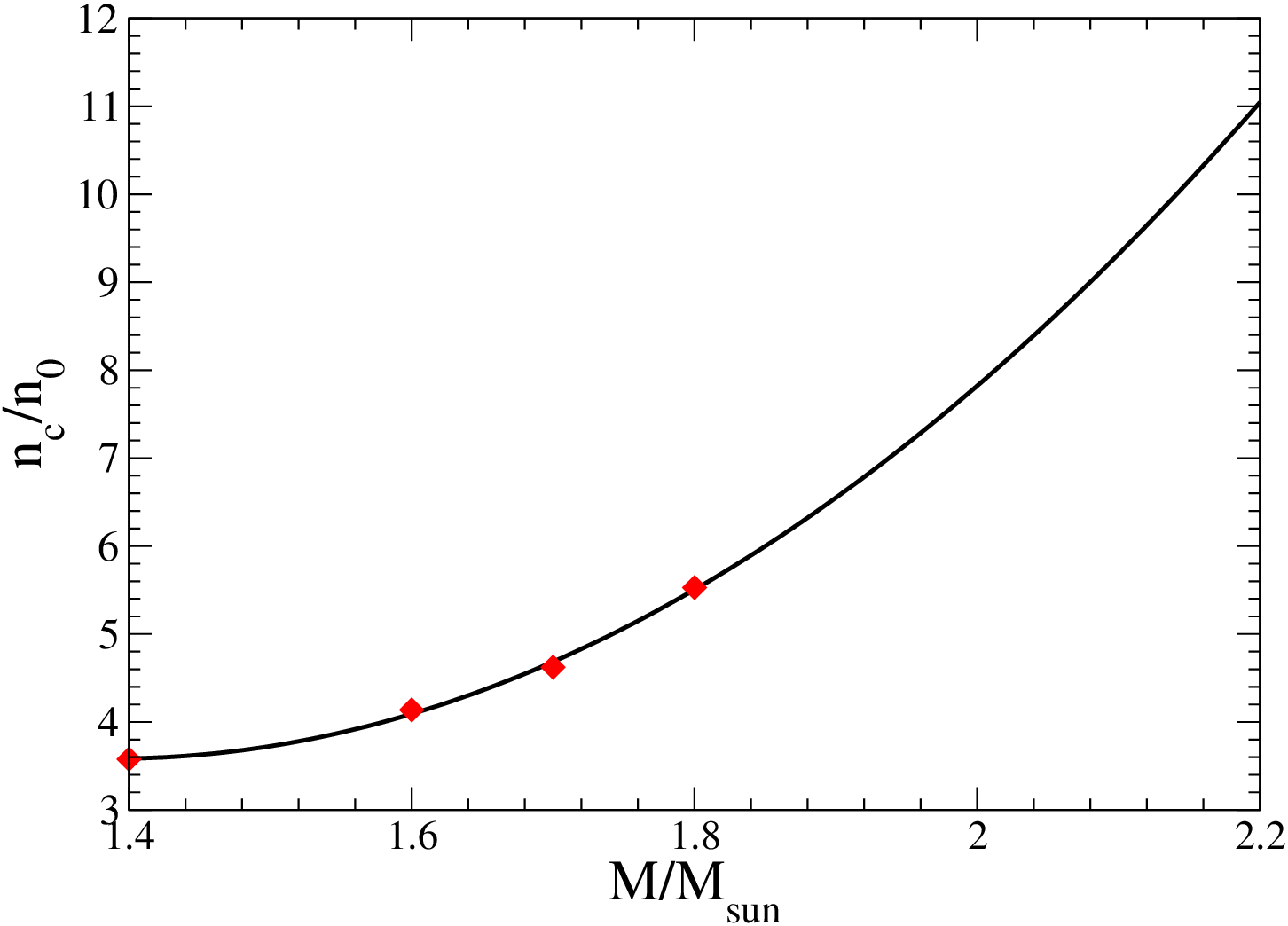}
 \caption{\label{rho_cxM}(Color online) Proton superconductivity depth
   as a function of neutron star mass. The solid line represents the
   best (2nd order) polynomial fit of the data.}
 \end{figure}

Figure \ref{pSC_prof} displays the critical temperatures of
superconducting protons for the neutron stars shown in
Fig.\ \ref{cool_R}. Their central densities can be read off from the
termination points of each curve.  Figure \ref{GAP} shows the density
dependence of the $^1S_0$ proton pairing gap for each stellar model.
This figure shows that the depth of the superconducting proton phase
increases with neutron star mass. This is a consequence of the direct
Urca process, whose fast cooling rate needs to be
suppressed by the superconducting protons. Otherwise the stellar model
would cool down much faster than observed for Cas A.  Aside from the
direct Urca process, superconducting protons are also essential for
controlling the changes in the stellar temperature caused by the
modified Urca process.  It if were not for the superconducting
protons, the core temperature would drop too quickly so that a
conflict with the emission of neutrinos from superfluid pair-breaking
neutrons would arise, again rendering agreement with the Cas A data
impossible. In summary, we conclude that, in agreement with
\cite{Yakovlev2011,Page2011}, proton superconductivity needs to be
present in the core of the neutron star in Cas A.  Agreement with the
observed temperatures of this object can however achieved for a range
of different proton superconducting profiles and critical
temperatures. The latter allows us to establish a connection between
the mass of a neutron star and the core penetration depth of
superconducting protons.  The result is shown in Fig. \ref{rho_cxM},
where we plot the density (core penetration depth) up to which
superconducting protons exist in the core of Cas A, considering a range
of possible stellar mass. The solid line represents a 2nd order
polynomial best fit to the numerical data. 
One sees that for lower stellar masses the curve flattens
considerably, approaching a penetration depth of $\rho_c / \rho_0
\simeq 3.75$ for a canonical neutron star mass of $1.4\, \msun$. The
flattening is consistent with the fact that for low stellar masses the
direct Urca process is highly suppressed due to low proton fractions
inside of such stars. Hence, the superconducting protons only need to
suppress the modified Urca process along with other less rapid stellar
cooling processes.
This relationship shown  in Fig.\ref{rho_cxM} can be used to either
predict the depth of the superconducting proton phase, or, conversely,
determine the mass of Cas A from a reliable calculation of the size of
the proton superconducting phase in superdense neutron star matter.

{\it Summary and Conclusions} In this letter, we have thoroughly
investigated the role of proton superconductivity for the cooling of
neutron stars, building on earlier work of
\cite{Yakovlev2011,Page2011}.  Our results confirm that
superconducting protons are required in order to explain the recent
temperature data observed for Cas A.  Furthermore our results agree
with the maximum critical temperature for superfluid $^3P_2$ neutron
established in \cite{Yakovlev2011,Page2011}.

In contrast to earlier work, however, we study a series of
phenomenological proton-pairing models to determine how deep into the
stellar core superconducting protons actually penetrate.  This allows
us to establish a heretofore unknown relationship between the mass of
the neutron star in Cas A and the penetration depth of the
superconducting proton phase. This relationship can be used to either
predict the depth of the superconducting proton phase, or, conversely,
determine the mass of Cas A from a reliable calculation of the size of
the proton superconducting phase in superdense neutron star matter.
We emphasize that the strategy outlined in this paper can be applied
to any other neutron star of similar age, whose temperature has been
reliably monitored over a several years period. High-mass neutron
stars, such as the recently discovered neutron stars J1614-2230 ($1.97
\pm 0.04\, \msun$) \cite{Demorest10} and J0348+0432 ($2.01 \pm 0.04 \,
\msun$) \cite{Antoniadis13}, appear particularly appealing as a significant
fraction of the protons in their cores may be superconducting.

This research clearly underlines the need for improved,
state-of-the-art calculations focusing on the role of protons (as
well as neutrons) in superdense neutron star matter.

{\bf Acknowledgment}
This material is based upon work supported by the
National Science Foundation under Grant No.\ 0854699.


\begin{thebibliography}{40}
\expandafter\ifx\csname natexlab\endcsname\relax\def\natexlab#1{#1}\fi
\expandafter\ifx\csname bibnamefont\endcsname\relax
  \def\bibnamefont#1{#1}\fi
\expandafter\ifx\csname bibfnamefont\endcsname\relax
  \def\bibfnamefont#1{#1}\fi
\expandafter\ifx\csname citenamefont\endcsname\relax
  \def\citenamefont#1{#1}\fi
\expandafter\ifx\csname url\endcsname\relax
  \def\url#1{\texttt{#1}}\fi
\expandafter\ifx\csname urlprefix\endcsname\relax\def\urlprefix{URL
}\fi
\providecommand{\bibinfo}[2]{#2}
\providecommand{\eprint}[2][]{\url{#2}}

\bibitem[{\citenamefont{Tsuruta and Cameron}(1965)}]{Tsuruta1965}
\bibinfo{author}{\bibfnamefont{S.}~\bibnamefont{Tsuruta}},
  \bibinfo{journal}{Nature} \textbf{\bibinfo{volume}{207}},
  \bibinfo{pages}{364} (\bibinfo{year}{1965}).

\bibitem[{\citenamefont{Maxwell}(1979)}]{Maxwell1979}
\bibinfo{author}{\bibfnamefont{O.~V.} \bibnamefont{Maxwell}},
  \bibinfo{journal}{The Astrophysical Journal}
\textbf{\bibinfo{volume}{231}},
  \bibinfo{pages}{201} (\bibinfo{year}{1979}).

\bibitem[{\citenamefont{Schaab et~al.}(1996)\citenamefont{Schaab,
Weber,
  Weigel, and Glendenning}}]{Schaab1996}
\bibinfo{author}{\bibfnamefont{C.}~\bibnamefont{Schaab}},
  \bibinfo{author}{\bibfnamefont{F.}~\bibnamefont{Weber}},
  \bibinfo{author}{\bibfnamefont{M.~K.}~\bibnamefont{Weigel}},
\bibnamefont{and}
  \bibinfo{author}{\bibfnamefont{N.~K.} \bibnamefont{Glendenning}},
  \bibinfo{journal}{Nuclear Phys A} \textbf{\bibinfo{volume}{605}},
  \bibinfo{pages}{531} (\bibinfo{year}{1996}).

\bibitem[{\citenamefont{Page et~al.}(2004)\citenamefont{Page,
Lattimer,
  Prakash, and Steiner}}]{Page2004}
\bibinfo{author}{\bibfnamefont{D.}~\bibnamefont{Page}},
  \bibinfo{author}{\bibfnamefont{J.}~\bibnamefont{Lattimer}},
  \bibinfo{author}{\bibfnamefont{M.}~\bibnamefont{Prakash}},
\bibnamefont{and}
  \bibinfo{author}{\bibfnamefont{A.~W.} \bibnamefont{Steiner}},
  \bibinfo{journal}{The Astrophysical Journal Supplement Series}
  \textbf{\bibinfo{volume}{155}}, \bibinfo{pages}{623}
(\bibinfo{year}{2004}).

\bibitem[{\citenamefont{Page et~al.}(2006)\citenamefont{Page, Geppert,
and
  Weber}}]{Page2006}
\bibinfo{author}{\bibfnamefont{D.}~\bibnamefont{Page}},
  \bibinfo{author}{\bibfnamefont{U.}~\bibnamefont{Geppert}},
\bibnamefont{and}
  \bibinfo{author}{\bibfnamefont{F.}~\bibnamefont{Weber}},
  \bibinfo{journal}{Nuclear Physics A} \textbf{\bibinfo{volume}{777}},
  \bibinfo{pages}{497} (\bibinfo{year}{2006}).

\bibitem[{\citenamefont{Page et~al.}(2009)\citenamefont{Page,
Lattimer,
  Prakash, and Steiner}}]{Page2009}
\bibinfo{author}{\bibfnamefont{D.}~\bibnamefont{Page}},
  \bibinfo{author}{\bibfnamefont{J.}~\bibnamefont{Lattimer}},
  \bibinfo{author}{\bibfnamefont{M.}~\bibnamefont{Prakash}},
\bibnamefont{and}
  \bibinfo{author}{\bibfnamefont{A.~W.} \bibnamefont{Steiner}},
  \bibinfo{journal}{The Astrophysical Journal}
\textbf{\bibinfo{volume}{707}},
  \bibinfo{pages}{1131} (\bibinfo{year}{2009}).

\bibitem[{\citenamefont{Blaschke et~al.}(2000)\citenamefont{Blaschke,
Klahn, and Voskresensky}}]{Blaschke2000}
\bibinfo{author}{\bibfnamefont{D.}~\bibnamefont{Blaschke}},
  \bibinfo{author}{\bibfnamefont{T.}~\bibnamefont{Klahn}},
\bibnamefont{and}
  \bibinfo{author}{\bibfnamefont{D.~N.}~\bibnamefont{Voskresensky}},
  \bibinfo{journal}{The Astrophysical Journal}
\textbf{\bibinfo{volume}{533}},
  \bibinfo{pages}{406} (\bibinfo{year}{2000}).

\bibitem[{\citenamefont{Grigorian
et~al.}(2005)\citenamefont{Grigorian,
  Blaschke, and Voskresensky}}]{Grigorian2005}
\bibinfo{author}{\bibfnamefont{H.}~\bibnamefont{Grigorian}},
  \bibinfo{author}{\bibfnamefont{D.}~\bibnamefont{Blaschke}},
\bibnamefont{and}
  \bibinfo{author}{\bibfnamefont{D.}~\bibnamefont{Voskresensky}},
  \bibinfo{journal}{Physical Review C} \textbf{\bibinfo{volume}{71}},
  \bibinfo{pages}{045801} (\bibinfo{year}{2005}).

\bibitem[{\citenamefont{Blaschke et~al.}(2006)\citenamefont{Blaschke,
  Voskresensky, and Grigorian}}]{Blaschke2006}
\bibinfo{author}{\bibfnamefont{D.}~\bibnamefont{Blaschke}},
  \bibinfo{author}{\bibfnamefont{D.}~\bibnamefont{Voskresensky}},
  \bibnamefont{and}
  \bibinfo{author}{\bibfnamefont{H.}~\bibnamefont{Grigorian}},
  \bibinfo{journal}{Nuclear Physics A} \textbf{\bibinfo{volume}{774}},
  \bibinfo{pages}{815} (\bibinfo{year}{2006}).

\bibitem[{\citenamefont{Page
et~al.}(2011{\natexlab{a}})\citenamefont{Page,
  Prakash, Lattimer, and Steiner}}]{Page2011}
\bibinfo{author}{\bibfnamefont{D.}~\bibnamefont{Page}},
  \bibinfo{author}{\bibfnamefont{M.}~\bibnamefont{Prakash}},
  \bibinfo{author}{\bibfnamefont{J.}~\bibnamefont{Lattimer}},
\bibnamefont{and}
  \bibinfo{author}{\bibfnamefont{A.~W.}~\bibnamefont{Steiner}},
  \bibinfo{journal}{Physical Review Letters}
\textbf{\bibinfo{volume}{106}},
  \bibinfo{pages}{081101} (\bibinfo{year}{2011}{\natexlab{a}}).

\bibitem[{\citenamefont{Yakovlev et~al.}(2011)\citenamefont{Yakovlev,
Ho,
  Shternin, Heinke, and Potekhin}}]{Yakovlev2011}
\bibinfo{author}{\bibfnamefont{D.~G.} \bibnamefont{Yakovlev}},
  \bibinfo{author}{\bibfnamefont{W.~C.~G.} \bibnamefont{Ho}},
  \bibinfo{author}{\bibfnamefont{P.~S.} \bibnamefont{Shternin}},
  \bibinfo{author}{\bibfnamefont{C.~O.} \bibnamefont{Heinke}},
  \bibnamefont{and} \bibinfo{author}{\bibfnamefont{A.~Y.}
  \bibnamefont{Potekhin}}, \bibinfo{journal}{Monthly Notices of the
Royal
  Astronomical Society} \textbf{\bibinfo{volume}{411}},
\bibinfo{pages}{1977}
  (\bibinfo{year}{2011}).


\bibitem[{\citenamefont{Schaab et~al.}(1997)\citenamefont{Schaab,
Voskresensky,
  Sedrakian, and Weber}}]{Voskresensky1997}
\bibinfo{author}{\bibfnamefont{C.}~\bibnamefont{Schaab}},
  \bibinfo{author}{\bibfnamefont{D.}~\bibnamefont{Voskresensky}},
  \bibinfo{author}{\bibfnamefont{A.}~\bibnamefont{Sedrakian}},
  \bibnamefont{and} \bibinfo{author}{\bibnamefont{F.}~\bibnamefont{Weber}},
  \bibinfo{journal}{Astronomy and Astrophysics}
\textbf{\bibinfo{volume}{604}},
  \bibinfo{pages}{591} (\bibinfo{year}{1997}).


\bibitem[{\citenamefont{Shternin et~al.}(2011)\citenamefont{Shternin,
Yakovlev,  Heinke, Ho, and Patnaude}}]{Shternin2011}
\bibinfo{author}{\bibfnamefont{P.~S.} \bibnamefont{Shternin}},
  \bibinfo{author}{\bibfnamefont{D.~G.} \bibnamefont{Yakovlev}},
  \bibinfo{author}{\bibfnamefont{C.~O.} \bibnamefont{Heinke}},
  \bibinfo{author}{\bibfnamefont{W.~C.~G.} \bibnamefont{Ho}},
\bibnamefont{and}
  \bibinfo{author}{\bibfnamefont{D.~J.} \bibnamefont{Patnaude}},
  \bibinfo{journal}{Monthly Notices of the Royal Astronomical Society:
Letters}
  \textbf{\bibinfo{volume}{412}}, \bibinfo{pages}{L108}
(\bibinfo{year}{2011}).

\bibitem[{\citenamefont{Shovkovy and Ellis}(2002)}]{Shovkovy2002}
\bibinfo{author}{\bibfnamefont{I.~A.}~\bibnamefont{Shovkovy}}
\bibnamefont{and}
  \bibinfo{author}{\bibfnamefont{P.~J.}~\bibnamefont{Ellis}},
  \bibinfo{journal}{Physical Review C} \textbf{\bibinfo{volume}{66}},
  \bibinfo{pages}{015802} (\bibinfo{year}{2002}).

\bibitem[{\citenamefont{Prakash et~al.}(1992)\citenamefont{Prakash,
Lattimer,
  and Pethick}}]{Prakash1992}
\bibinfo{author}{\bibfnamefont{M.}~\bibnamefont{Prakash}},
  \bibinfo{author}{\bibfnamefont{J.}~\bibnamefont{Lattimer}},
\bibnamefont{and}
  \bibinfo{author}{\bibfnamefont{C.}~\bibnamefont{Pethick}},
  \bibinfo{journal}{Astrophysical Journal}
\textbf{\bibinfo{volume}{390}},
  \bibinfo{pages}{L77} (\bibinfo{year}{1992}).

\bibitem[{\citenamefont{Lattimer et~al.}(1994)\citenamefont{Lattimer,
van
  Riper, Prakash, and Prakash}}]{Lattimer1994}
\bibinfo{author}{\bibfnamefont{J.~M.} \bibnamefont{Lattimer}},
  \bibinfo{author}{\bibfnamefont{K.~A.} \bibnamefont{van Riper}},
  \bibinfo{author}{\bibfnamefont{M.}~\bibnamefont{Prakash}},
\bibnamefont{and}
  \bibinfo{author}{\bibfnamefont{M.}~\bibnamefont{Prakash}},
  \bibinfo{journal}{The Astrophysical Journal}
\textbf{\bibinfo{volume}{425}},
  \bibinfo{pages}{802} (\bibinfo{year}{1994}).

\bibitem[{\citenamefont{Gusakov et~al.}(2005)\citenamefont{Gusakov,
Kaminker,
  Yakovlev, and Gnedin}}]{Gusakov2005}
\bibinfo{author}{\bibfnamefont{M.~E.} \bibnamefont{Gusakov}},
  \bibinfo{author}{\bibfnamefont{A.~D.} \bibnamefont{Kaminker}},
  \bibinfo{author}{\bibfnamefont{D.~G.} \bibnamefont{Yakovlev}},
  \bibnamefont{and} \bibinfo{author}{\bibfnamefont{O.~Y.}
  \bibnamefont{Gnedin}}, \bibinfo{journal}{Monthly Notices of the
Royal
  Astronomical Society} \textbf{\bibinfo{volume}{363}},
\bibinfo{pages}{555}
  (\bibinfo{year}{2005}).

\bibitem[{\citenamefont{Horvath et~al.}(1991)\citenamefont{Horvath,
Benvenuto,
  and Vucetich}}]{Horvath1991}
\bibinfo{author}{\bibfnamefont{J.}~\bibnamefont{Horvath}},
  \bibinfo{author}{\bibfnamefont{O.}~\bibnamefont{Benvenuto}},
  \bibnamefont{and}
\bibinfo{author}{\bibfnamefont{H.}~\bibnamefont{Vucetich}},
  \bibinfo{journal}{Physical Review D} \textbf{\bibinfo{volume}{44}},
  \bibinfo{pages}{3797} (\bibinfo{year}{1991}).

\bibitem[{\citenamefont{Weber et~al.}(2009)\citenamefont{Weber,
Negreiros, and
  Rosenfield}}]{Weber2007b}
\bibinfo{author}{\bibfnamefont{F.}~\bibnamefont{Weber}},
  \bibinfo{author}{\bibfnamefont{R.}~\bibnamefont{Negreiros}},
  \bibnamefont{and}
  \bibinfo{author}{\bibfnamefont{P.}~\bibnamefont{Rosenfield}},
  \bibinfo{journal}{Astrophysics and Space Science Library}
  \textbf{\bibinfo{volume}{357}}, \bibinfo{pages}{213}
(\bibinfo{year}{2009}).

\bibitem[{\citenamefont{Negreiros
et~al.}(2010)\citenamefont{Negreiros,
  Dexheimer, and Schramm}}]{Negreiros2010}
\bibinfo{author}{\bibfnamefont{R.}~\bibnamefont{Negreiros}},
  \bibinfo{author}{\bibfnamefont{V.~A.} \bibnamefont{Dexheimer}},
  \bibnamefont{and}
\bibinfo{author}{\bibfnamefont{S.}~\bibnamefont{Schramm}},
  \bibinfo{journal}{Physical Review C} \textbf{\bibinfo{volume}{82}},
  \bibinfo{pages}{035803} (\bibinfo{year}{2010}).

\bibitem[{\citenamefont{Niebergal
et~al.}(2010)\citenamefont{Niebergal, Ouyed,
  Negreiros, and Weber}}]{Niebergal}
\bibinfo{author}{\bibfnamefont{B.}~\bibnamefont{Niebergal}},
  \bibinfo{author}{\bibfnamefont{R.}~\bibnamefont{Ouyed}},
  \bibinfo{author}{\bibfnamefont{R.}~\bibnamefont{Negreiros}},
  \bibnamefont{and}
\bibinfo{author}{\bibfnamefont{F.}~\bibnamefont{Weber}},
  \bibinfo{journal}{Physical Review D} \textbf{\bibinfo{volume}{81}},
  \bibinfo{pages}{043005} (\bibinfo{year}{2010}).

\bibitem[{\citenamefont{Alford et~al.}(2005)\citenamefont{Alford,
Jotwani,
  Kouvaris, Kundu, and Rajagopal}}]{Alford2005a}
\bibinfo{author}{\bibfnamefont{M.}~\bibnamefont{Alford}},
  \bibinfo{author}{\bibfnamefont{P.}~\bibnamefont{Jotwani}},
  \bibinfo{author}{\bibfnamefont{C.}~\bibnamefont{Kouvaris}},
  \bibinfo{author}{\bibfnamefont{J.}~\bibnamefont{Kundu}},
\bibnamefont{and}
  \bibinfo{author}{\bibfnamefont{K.}~\bibnamefont{Rajagopal}},
  \bibinfo{journal}{Physical Review D} \textbf{\bibinfo{volume}{71}},
  \bibinfo{pages}{114011} (\bibinfo{year}{2005}).


 \bibitem{Negreiros2012a} R. Negreiros, S. Schramm, and F. Weber,
   Physical Review. D 85, 104019 (2012).
%
  
\bibitem[{\citenamefont{Heinke and Ho}(2010)}]{Heinke2010}
\bibinfo{author}{\bibfnamefont{C.~O.} \bibnamefont{Heinke}}
\bibnamefont{and}
  \bibinfo{author}{\bibfnamefont{W.~C.~G.} \bibnamefont{Ho}},
  \bibinfo{journal}{The Astrophysical Journal}
\textbf{\bibinfo{volume}{719}},
  \bibinfo{pages}{L167} (\bibinfo{year}{2010}).


\bibitem{Blaschke2011} D. Blaschke, H. Grigorian, D. N. Voskresensky,
  and F. Weber, Physical Review C 85, 022802 (2012).
 
\bibitem{Negreiros2013} R. Negreiros, S. Schramm, and F. Weber,
   Physics Letters B 718, 1176 (2013).
 
 
 \bibitem[Page et al.(2011)]{Page2011b} D.~P. Page, M. Prakash, 
J. Lattimer, \& A. Steiner,  Brazilian Workshop on Nuclear Physics (XXXIV
BWNP) (2011).



\bibitem[{Akmal et~al.(1998)Akmal, Pandharipande, and Ravenhall}]{Akmal1998}
\bibinfo{author}{A.~Akmal}, \bibinfo{author}{V.~R. Pandharipande},
  \bibinfo{author}{D.~G. Ravenhall},
\newblock \bibinfo{journal}{Physical Review C} \bibinfo{volume}{58},
   \bibinfo{pages}{1804} (\bibinfo{year}{1998}).


\bibitem[{Heiselberg and Hjorth-Jensen(2000)}]{Heiselberg2000}
\bibinfo{author}{H.~Heiselberg}, \bibinfo{author}{M.~Hjorth-Jensen},
\newblock \bibinfo{journal}{Physics Reports} \bibinfo{volume}{328},
  \bibinfo{pages}{237} (\bibinfo{year}{2000}).


\bibitem[{\citenamefont{Weber}(1999)}]{Weber}
\bibinfo{author}{\bibfnamefont{F.}~\bibnamefont{Weber}},
  {\it {Pulsars as astrophysical laboratories for nuclear and
  particle physics}} (\bibinfo{publisher}{Institute of Physics},
  \bibinfo{address}{Bristol}, \bibinfo{year}{1999}).



\bibitem[{\citenamefont{Glendenning}(2000)}]{Glendenning2000}
\bibinfo{author}{\bibfnamefont{N.~K.} \bibnamefont{Glendenning}},
  {\bibinfo{title}{ \it {Compact stars: nuclear physics, particle
physics, and
  general relativity}}} (\bibinfo{publisher}{Springer},
\bibinfo{year}{2000}),
  \bibinfo{edition}{2nd} ed.


\bibitem[{\citenamefont{Komatsu et~al.}(1989)\citenamefont{Komatsu,
Eriguchi,
  and Hachisu}}]{Komatsu1989}
\bibinfo{author}{\bibfnamefont{H.}~\bibnamefont{Komatsu}},
  \bibinfo{author}{\bibfnamefont{Y.}~\bibnamefont{Eriguchi}},
\bibnamefont{and}
  \bibinfo{author}{\bibfnamefont{I.}~\bibnamefont{Hachisu}},
  \bibinfo{journal}{Royal Astronomical Society, Monthly Notices }
\textbf{\bibinfo{volume}{237}},
\bibinfo{pages}{355} (\bibinfo{year}{1989}).

\bibitem[{\citenamefont{Cook et~al.}(1992)\citenamefont{Cook, Shapiro,
and
  Teukolsky}}]{Cook1992}
\bibinfo{author}{\bibfnamefont{G.~B.} \bibnamefont{Cook}},
  \bibinfo{author}{\bibfnamefont{S.~L.} \bibnamefont{Shapiro}},
  \bibnamefont{and} \bibinfo{author}{\bibfnamefont{S.~A.}
  \bibnamefont{Teukolsky}}, \bibinfo{journal}{The Astrophysical
Journal}
  \textbf{\bibinfo{volume}{398}}, \bibinfo{pages}{203}
(\bibinfo{year}{1992}).

\bibitem[{\citenamefont{Stergioulas and
Friedman}(1995)}]{Stergioulas1995}
\bibinfo{author}{\bibfnamefont{N.}~\bibnamefont{Stergioulas}}
\bibnamefont{and}
  \bibinfo{author}{\bibfnamefont{J.~L.} \bibnamefont{Friedman}},
  \bibinfo{journal}{The Astrophysical Journal}
\textbf{\bibinfo{volume}{444}},
  \bibinfo{pages}{306} (\bibinfo{year}{1995}).

\bibitem[{\citenamefont{Glendenning}(1989)}]{Glendenning1989}
\bibinfo{author}{\bibfnamefont{N.~K.} \bibnamefont{Glendenning}},
  \bibinfo{journal}{Nuclear Physics A} \textbf{\bibinfo{volume}{493}},
  \bibinfo{pages}{521} (\bibinfo{year}{1989}).

\bibitem{Schaab1998} C. Schaab and M. K. Weigel, Astronomy and
  Astrophysics {\bf 336}, L13 (1998).

\bibitem[{\citenamefont{Stejner et~al.}(2009)\citenamefont{Stejner,
Weber, and
  Madsen}}]{Stejner2009}
\bibinfo{author}{\bibfnamefont{M.}~\bibnamefont{Stejner}},
  \bibinfo{author}{\bibfnamefont{F.}~\bibnamefont{Weber}},
\bibnamefont{and}
  \bibinfo{author}{\bibfnamefont{J.}~\bibnamefont{Madsen}},
  \bibinfo{journal}{The Astrophysical Journal}
\textbf{\bibinfo{volume}{694}},
  \bibinfo{pages}{1019} (\bibinfo{year}{2009}).


\bibitem{Cheng2004}
C. Y. Hui, and K. S. Cheng, The  Astrophysical Journal {\bf 608},
935 (2004).

\bibitem[{\citenamefont{Pons et~al.}(2009)\citenamefont{Pons,
Miralles, and
  Geppert}}]{Pons2009}
\bibinfo{author}{\bibfnamefont{J.~A.} \bibnamefont{Pons}},
  \bibinfo{author}{\bibfnamefont{J.~A.} \bibnamefont{Miralles}},
  \bibnamefont{and}
\bibinfo{author}{\bibfnamefont{U.}~\bibnamefont{Geppert}},
  \bibinfo{journal}{Astronomy and Astrophysics}
\textbf{\bibinfo{volume}{496}},
  \bibinfo{pages}{207} (\bibinfo{year}{2009}).


\bibitem{Aguilera2008}
D.~N. Aguilera, J.~A. Pons, and J.~A. Miralles, Astronomy and
Astrophysics {\bf 486}, 255 (2008).

\bibitem[{\citenamefont{Lattimer et~al.}(1991)\citenamefont{Lattimer,
Pethick,
  Prakash, and Haensel}}]{Lattimer1991}
\bibinfo{author}{\bibfnamefont{J.~M}~\bibnamefont{Lattimer}},
  \bibinfo{author}{\bibfnamefont{C.}~\bibnamefont{Pethick}},
  \bibinfo{author}{\bibfnamefont{M.}~\bibnamefont{Prakash}},
\bibnamefont{and}
  \bibinfo{author}{\bibfnamefont{P.}~\bibnamefont{Haensel}},
  \bibinfo{journal}{Physical Review Letters}
\textbf{\bibinfo{volume}{66}},
  \bibinfo{pages}{2701} (\bibinfo{year}{1991}).

\bibitem[{\citenamefont{Levenfish and Yakovlev}(1994)}]{Levenfish1994}
\bibinfo{author}{\bibfnamefont{K.~P.} \bibnamefont{Levenfish}}
  \bibnamefont{and} \bibinfo{author}{\bibfnamefont{D.~G.}
  \bibnamefont{Yakovlev}}, \bibinfo{journal}{Astronomy Letters}
  \textbf{\bibinfo{volume}{20}}, \bibinfo{pages}{43}
(\bibinfo{year}{1994}).


\bibitem[{\citenamefont{Gnedin et~al.}(2001)\citenamefont{Gnedin,
Yakovlev, and
  Potekhin}}]{Gnedin2001}
\bibinfo{author}{\bibfnamefont{O.~Y.} \bibnamefont{Gnedin}},
  \bibinfo{author}{\bibfnamefont{D.~G.} \bibnamefont{Yakovlev}},
  \bibnamefont{and} \bibinfo{author}{\bibfnamefont{A.~Y.}
  \bibnamefont{Potekhin}}, \bibinfo{journal}{Monthly Notices of the
Royal
  Astronomical Society} \textbf{\bibinfo{volume}{324}},
\bibinfo{pages}{725}
  (\bibinfo{year}{2001}).

\bibitem[{\citenamefont{Yakovlev et~al.}(2001)\citenamefont{Yakovlev,
Kaminker,
  Gnedin, and Haensel}}]{Yakovlev2001a}
\bibinfo{author}{\bibfnamefont{D.~G.} \bibnamefont{Yakovlev}},
  \bibinfo{author}{\bibfnamefont{A.~D.} \bibnamefont{Kaminker}},
  \bibinfo{author}{\bibfnamefont{O.~Y.} \bibnamefont{Gnedin}},
  \bibnamefont{and}
\bibinfo{author}{\bibfnamefont{P.}~\bibnamefont{Haensel}},
  \bibinfo{journal}{Physics Reports} \textbf{\bibinfo{volume}{354}},
  \bibinfo{pages}{1} (\bibinfo{year}{2001}).

\bibitem[{\citenamefont{Flowers and Itoh}(1981)}]{Flowers1981}
\bibinfo{author}{\bibfnamefont{E.}~\bibnamefont{Flowers}}
\bibnamefont{and}
  \bibinfo{author}{\bibfnamefont{N.}~\bibnamefont{Itoh}},
\bibinfo{journal}{The
  Astrophysical Journal} \textbf{\bibinfo{volume}{250}},
\bibinfo{pages}{750}
  (\bibinfo{year}{1981}).

\bibitem{Ott2006}
C. Ott, A. Burrows, T.A. Thompson, E. Livne, and R.
Walder, The Astrophysical Journal Supplement Series {\bf 164}, 130
(2006).

\bibitem{Kaspi2006}
C. A. Faucher-Giguere and V.M. Kaspi, The Astrophysical Journal {\bf
643}, 332 (2006).


\bibitem{hessels06:a} J. W. Hessels, S. M. Ransom, I. H. Stairs,
  P. C. Freire, V. M.  Kaspi, and F. Camilo, Science {\bf 311}
  (No.\ 5769), 1901 (2006).

 
\bibitem{backer82:a} D. C. Backer, S. R. Kulkarni, C. Heiles,
  M. M. Davis, and W. M. Goss, Nature {\bf 300} (1982) 615.

\bibitem{Lowell2012}
A.W. Lowell, J.A. Tomsick, C.O. Heinke, A. Bodaghee,
S.E. Boggs, P. Kaaret, S. Chaty, J. Rodriguez, and
R. Walter, {\tt arXiv:1202.1531v1 [astro-ph.HE]} (2012).

\bibitem[{\citenamefont{Gudmundsson
et~al.}(1983)\citenamefont{Gudmundsson,
  Pethick, and Epstein}}]{Gudmundsson1983}
\bibinfo{author}{\bibfnamefont{E.~H.} \bibnamefont{Gudmundsson}},
  \bibinfo{author}{\bibfnamefont{C.~J.} \bibnamefont{Pethick}},
  \bibnamefont{and} \bibinfo{author}{\bibfnamefont{R.~I.}
  \bibnamefont{Epstein}}, \bibinfo{journal}{The Astrophysical Journal}
  \textbf{\bibinfo{volume}{272}}, \bibinfo{pages}{286}
(\bibinfo{year}{1983}).

\bibitem[{\citenamefont{Potekhin et~al.}(1997)\citenamefont{Potekhin,
Chabrier,
  and Yakovlev}}]{Potekhin1997}
\bibinfo{author}{\bibfnamefont{A.~Y.} \bibnamefont{Potekhin}},
  \bibinfo{author}{\bibfnamefont{G.}~\bibnamefont{Chabrier}},
\bibnamefont{and}
  \bibinfo{author}{\bibfnamefont{D.~G.} \bibnamefont{Yakovlev}},
  \bibinfo{journal}{Astronomy and Astrophysics}
\textbf{\bibinfo{volume}{323}},
  \bibinfo{pages}{415} (\bibinfo{year}{1997}).

\bibitem{Demorest10} P. B. Demorest, T. Pennucci, S. M. Ransom,
  M. S. E. Roberts, and J. W. T.  Hessels, Nature {\bf 467}, 1081 (2010).

\bibitem{Antoniadis13} J. Antoniadis {\it et al.},
   Science {\bf 340}, no.\ 6131 (2013).

\end{thebibliography}

\end{document}